\documentclass[preprint,aps]{revtex4}
\usepackage{graphics}
\usepackage{graphicx}
\usepackage{dcolumn}
\usepackage{bm}

\begin{document}
\title{Spin polarization versus lifetime effects at point contacts between superconducting niobium and normal metals}

\author{Elina Tuuli$^{1-3}$ and Kurt Gloos$^{1,3}$}

\address{$^1$ Wihuri Physical Laboratory, Department of Physics and Astronomy, FIN-20014 University of Turku, Finland}
\address{$^2$ Graduate School of Materials Research (GSMR), FIN-20500 Turku, Finland}
\address{$^3$ Turku University Centre for Materials and Surfaces (MatSurf), FIN-20014 Turku, Finland}


\begin{abstract}
Point-contact Andreev reflection spectroscopy is used to measure the spin polarization of metals but analysis of the spectra has encountered a number of serious challenges, one of which is the difficulty to distinguish the effects of spin polarization from those of the finite lifetime of Cooper pairs. We have recently confirmed the polarization-lifetime ambiguity for Nb-Co and Nb-Cu contacts and suggested to use Fermi surface mismatch, the normal reflection due to the difference of Fermi wave vectors of the two electrodes, to solve this dilemma. Here we present further experiments on contacts between superconducting Nb and the ferromagnets Fe and Ni as well as the noble metals Ag and Pt that support our previous results. Our data indicate that the Nb - normal metal interfaces have a transparency of up to about 80\% and a small, if not negligible, spin polarization.
\end{abstract}

\maketitle 

\section{Introduction}
With the growing interest in spintronics, measuring the spin polarization $P$ of a metal has become an important task~\cite{Prinz}. Andreev reflection spectroscopy of point contacts has been suggested as an effective and versatile technique for measuring the polarization of metals in contact with a superconductor~\cite{Jong,Soulen}. To extract the polarization, one analyses the measured differential resistance spectra with the help of a modified BTK model~\cite{BTK,Strijkers,Woods}. This method has been applied to a wide range of materials, delivering converging values for the spin polarization. The extracted $P$ has a linear or parabolic dependence on the relative strength of the interface barrier $Z$, and the true value of the polarization $P$ is measured in the $Z = 0$ limit. Unfortunately, this procedure has also encountered problems, including inadequately understood additional spectral features~\cite{Baltz} and, most of all, the degeneracy of the results of the various BTK type models~\cite{Bugoslavsky,Chalsani}. This applies especially when the polarization is low.

We have recently demonstrated how difficult it is to distinguish the effects of polarization and finite lifetime of Cooper pairs in the spectra of Nb-Co and Nb-Cu contacts~\cite{oma}. The finite lifetime of the Cooper pairs has been used to describe the spectra of non-magnetic contacts~\cite{Plecenik}, but with few exceptions~\cite{Bugoslavsky,Chalsani,Mukhopadhyay} it has not generally been applied to contacts with magnetic metals even though one should expect strong pair breaking in that case. Since spectra of both magnets and non-magnets can often be fitted rather well with the two models, we have suggested to use the $Z$ parameter as a distinguishing criterion between them. In addition to Nb - Co and Nb - Cu data, we present here measurements of Nb in contact with the ferromagnets Fe and Ni as well as the non-magnets Ag and Pt that support our previous results.

\section{Experimental details and results}

We formed the point contacts between a Nb wire and a normal metal wire with typical diameters of 0.25 mm using the shear method~\cite{kirja}. Their differential resistance was recorded in the standard four-wire scheme. Most of the measurements were carried out at 4.2 K in liquid helium and a smaller number in vacuum down to 0.7 K. The normal state resistance of the analysed contacts varied between 1 and 100 Ohm.

\begin{figure}[h]
	\includegraphics[width= \textwidth]{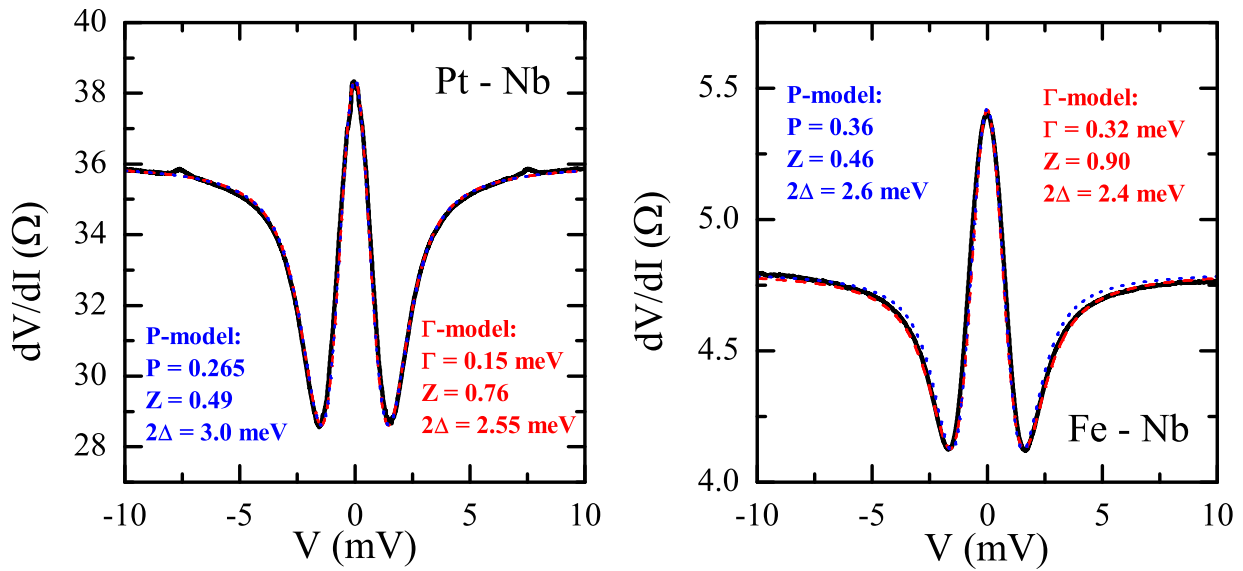}
	\caption{\label{fitf} Representative examples of Pt - Nb (left) and Fe - Nb (right) spectra measured in liquid He at 4.2 K. Black lines are the measured spectra, blue dotted lines the polarization fit, and red dashed line the lifetime fit.}
	\end{figure}

\begin{figure}
\begin{center}
\includegraphics[width=\textwidth]{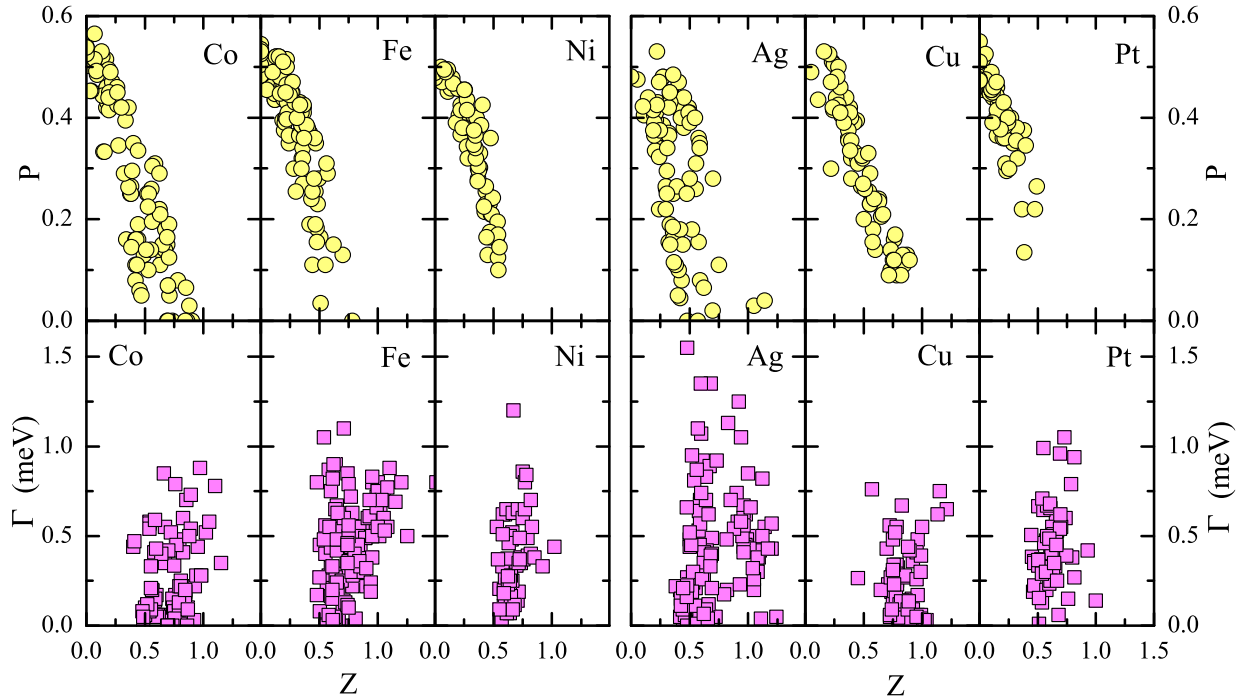}
\end{center}
\caption{\label{tulokset} The $P(Z)$ (top) and the $\Gamma(Z)$ (bottom) distributions of the investigated Nb - normal metal combinations.}
\end{figure}

Only spectra that showed the typical Andreev reflection double minimum structure were analysed with two different modified BTK models: Strijkers' model~\cite{Strijkers} that uses spin polarization to explain the spectral features and Plecen\'{i}k's model~\cite{Plecenik} as alternative description using the finite Cooper pair lifetime. Figure~\ref{fitf} shows that both models fit the measured spectra well. The parabolic $P(Z)$ dependence of ferromagnetic Co, Fe and Ni extracted using the polarization model agreed well with those measured by others~\cite{Strijkers,Baltz}. However, the same analysis of the non-magnetic metals implies a similar polarization also of Ag, Cu and Pt (Figure~\ref{tulokset}). The analysis of the contacts with the lifetime model results in the $\Gamma(Z)$ dependence also shown in Figure~\ref{tulokset}.

\section{Discussion}

It is obvious that the two models can not both be valid at the same time. Therefore we need to distinguish the two different interpretations from each other and find the one that is more plausible. The $Z$ dependence seems to be the most striking difference between the results of the two models~\cite{oma}. Figure~\ref{tulokset} shows that the polarization model leads to a parabolic $P(Z)$ dependence while the lifetime model gives $\Gamma$ at an almost constant $Z$. 

The $Z$ parameter can be expressed as $Z = \sqrt{ Z_b^2 + Z_0^2}$ where $Z_b$ represents the strength of a real tunneling barrier and $Z_0$ the reflections caused by the mismatch of the electronic structures on both sides of the interface~\cite{Blonder1983}. A simple estimate for the latter is $Z_0 = |1-r|/2\sqrt{r}$ with the ratio of the Fermi velocities of the two electrodes $r = v_{F_1}/v_{F_2}$~\cite{Blonder1983}. This approach neglects much of the complexity of the interfaces. However, obtaining even such an estimate is still not trivial because the Fermi velocities are difficult to determine experimentally, especially for the ferromagnets. Estimates for $Z_0$ range from less than 0.1 up to 0.4 for these metal combinations~\cite{oma}.



The $Z$ parameter is related to the transmission coefficient $\tau = 1/\sqrt{1+Z^2}$ of an interface. Transmission through interfaces has been investigated using the superconducting proximity effect (PE)~\cite{Cirillo2004,Tesauro2005,Attanasio2006,Kushnir2009} and the current perpendicular to plane magnetoresistance (CPP-MR)~\cite{Stiles2000,Xu2006,Sharma2007,Park2000}. In these experiments oxide barriers can be excluded because of the preparation method, and the transmission coefficient should therefore represent the contribution from Fermi surface mismatch only. The obtained transmission coefficients for interfaces between Nb and the non-magnetic normal metals are close to 0.3 which corresponds to $Z \approx 1.5$. The proximity effect studies indicate significantly smaller $\tau$ values for the ferromagnets in contact with Nb while crude estimates based on the free electron model and the CPP-MR data~\cite{Sharma2007,Park2000} result in transmission coefficients smaller than 0.5.


Our results in Figure~\ref{tulokset} derived from the polarization model indicate a very high transmission. Some of the contacts have even $Z \approx 0$ which means a nearly perfect transmission with $\tau \approx 1$. The lifetime model on the other hand indicates a maximum transmission coefficient of approximately 0.8 which is determined by the onset of the $\Gamma(Z)$ data at $Z \approx 0.5$. Hence, results of the lifetime model agree much better with the transmission coefficients obtained by the above mentioned different experimental methods than those of the polarization model.

Neither of the two models leads to a large difference between the ferromagnetic and non-magnetic metals. The $P(Z)$ dependency and the amount of pair breaking are almost identical for all of the investigated normal metals in contact with superconducting Nb. This is unexpected because spin polarization should reduce the amount of Andreev reflection and increase pair breaking at interfaces with ferromagnets more than with non-magnets. This should have a measurable effect on the spectra - why can it not be observed?


\section{Conclusions}

We have measured superconducting Nb in contact with both ferromagnetic and non-magnetic normal metals and analysed the resulting spectra with the mutually exclusive polarization and lifetime models. There are no clear differences between the magnetically ordered and non-magnetic metals in either of the two models or the measured spectra. However, the lifetime model fits better the experimental results obtained by other methods. This would indicate that point-contact Andreev reflection spectroscopy, at least with superconducting Nb, does possibly not deliver the true spin polarization of the normal metal.

\section{Acknowledgements}
We thank the Jenny and Antti Wihuri Foundation and the Finnish Concordia Fund for financial support.

\end{document}